# Mobility and sheet charge in high electron mobility transistor quantum wells from photon-induced transconductance

Yury Turkulets, and Ilan Shalish*

*Ben Gurion University of the Negev, Beer Sheva, Israel*

When a high electron mobility transistor is illuminated, the absorbed photons excite electron-hole pairs. The generated pairs are separated by the built in field in such a way that the electrons end up in the quantum well generating a photocurrent, while together with the holes that are swept towards the gate they generate a surface photovoltage. Here, we define photon-induced transconductance as the ratio between the surface photovoltage and the 2DEG photocurrent under identical illumination conditions. We show that this ratio directly yields the channel mobility and the 2DEG sheet charge density. The photocurrent and photovoltage may vary with the wavelength of the exciting photons. We examine and analyze optical spectra of this photon-induced transconductance obtained from an AlGaN/GaN heterostructure for a range of photon energies showing that the mobility is obtained only for excitation at photon energies above the wide bandgap energy. The method offers an optical alternative to Hall effect and to field effect mobility. Unlike Hall effect, it may be measured in the transistor itself. The only alternative that can measure mobility in the transistor itself measures field effect mobility, while the proposed method mesures the same conductivity mobility as measured by Hall effect.

The main application of the high electron mobility transistor (HEMT) today is radio frequency (RF) and power switching.[1,2] Its main advantage over other field effect transistors is in its low channel resistance that results from its high electron mobility and high sheet charge concentration.[3] Optimization of these parameters has been a central design issue that relies on characterization, with Hall effect being the prevalent method.[4,5] In most cases, the measurements are carried out either on bare wafers,[6,7] or on a dedicated van der Pauw test structures1 but seldom on the transistor itself.[8] However, fabrication processes often intentionally alter the sheet charge, and, as a result, the mobility changes as well. For example in GaN-based devices, certain passivation coatings are used to increase the density of surface state, which in turn, increase the density of charge in the 2-dimensional electron gas (2DEG).[9,10] These changes may vary among different transistors on the same wafer. Measuring mobility in the transistor has so far been limited to measuring the *field-effect mobility* that has been shown to be prone to errors especially in devices having extreme electron mobilities.[11]

In this paper, we propose an electro-optical method to characterize electron mobility and sheet charge density of the 2DEG in HEMT structures within the actual device, avoiding the field effect which has known errors in estimation of the mobility.[11,12,13,14] We demonstrate the method using an AlGaN/GaN heterostructure. The method requires a light source capable of emitting above the wide bandgap energy, a Kelvin probe, and two Ohmic contacts, such as the source and the drain. The Kelvin probe is used for measuring the surface voltage.[15,16] The two Ohmic contacts are used for applying voltage and measuring current. Both types of measurements are carried out in the dark as well as under illumination. Subtracting the dark from light values, we get the *surface photovoltage* and the *channel photocurrent* for the specific photon energy used.

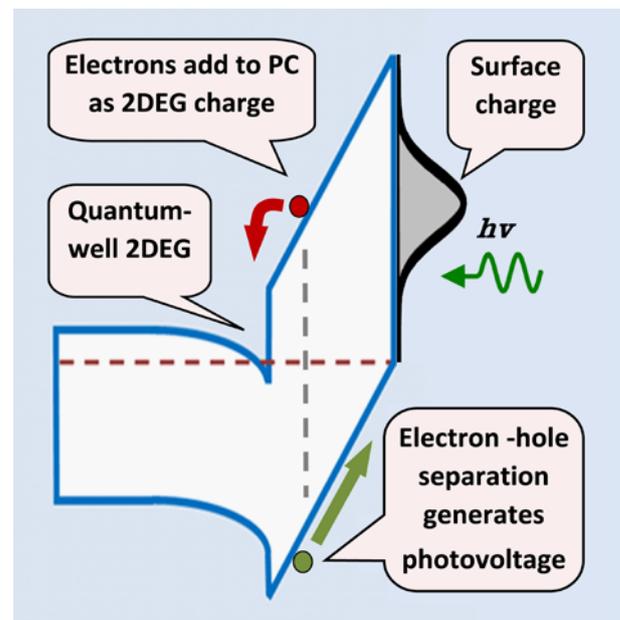

**Fig. 1** – Schematic band diagram of an AlGaN/GaN heterosctructure. Optically generated electron-hole pairs in the AlGaN layer are separated by the built-in field. Electrons are emitted into the quantum-well, adding to the 2DEG sheet charge. Holes are swept to the surface generating surface photovoltage.

To describe the method, we will refer to the AlGaN/GaN structure, but the same applies to any other heterojunction. Figure 1 shows a schematic band diagram of the AlGaN/GaN heterojunction. The top surface on which the illumination impinges is the AlGaN surface. It has surface states and surface charge trapped in these states. Above-bandgap illumination generates electron-hole pairs. Pairs generated within the AlGaN layer are separated by the built-in electric field in the layer. They are swept to the two sides of the AlGaN layer forming a photovoltage in the layer. Electrons swept into the triangular quantum well at the AlGaN/GaN interface increase the 2DEG charge density and give rise to the observed *channel photoconuctivity*. The channel photoconductivity is a product of the number density of the optically generated excess electrons, $\Delta n_s$, and the 2DEG electron mobility, $\mu$. Using Ohm's law, we can express the relation between the observed photocurrent, $\Delta I_D$, and the voltage, $V_{DS}$, applied between the source and the drain:

$$\Delta I_D = V_{DS} q \mu \Delta n_s \frac{W}{L} \qquad (1)$$

Where $q$ is the electron charge, and $W$ and $L$ are the width and length of the channel, respectively. Note that since the resistor is a quantum well, its resistivity is defined using sheet charge density rather than the common 3D charge density: $\frac{1}{R} = \frac{\sigma A}{L} = q\mu \frac{\Delta n}{d} \cdot \frac{dW}{L}$, where $d$, the thickness of the well, cancels out. Equation 1 has two unknowns, the mobility, and the optically generated 2DEG charge, which come as a product. To separate this product to its constituents, one has to know the value of one of them. In the commonly used Hall effect method, the charge density is obtained from the Hall voltage.[17] In our proposed method, we obtain it from the *surface photovoltage*.

The AlGaN layer may be viewed as a parallel-plate capacitor, to which plates two equal sheet charges of opposite signs were added (the 2DEG on the one side and the surface states on the other). The induced change in the sheet charge should be equal to the change in the electrical displacement across the AlGaN capacitor:

$$q\Delta n_s = \varepsilon_{AlGaN} \frac{\Delta \Phi_{BB}}{t_{AlGaN}} \qquad (2)$$

Where $\varepsilon_{AlGaN}$ is the dielectric constant of the AlGaN, $t_{AlGaN}$ is the thickness of the AlGaN layer, and $\Delta\Phi_{BB}$ is the surface photovoltage, or the illumination-induced change in the AlGaN band-bending. Since $\Delta\Phi_{BB}$ is the illumination-induced change in the voltage between the gate and the source, and $\Delta I_D$ is the illumination-induced change in the drain current, their ratio is an *illumination-induced transconductance*:

$$g_m^{OPT} = \frac{\Delta I_D}{\Delta \Phi_{BB}} = \varepsilon_{AlGaN} V_{DS} \mu \frac{W}{L t_{AlGaN}} \qquad (3)$$

In Eq. 3, all the parameters are known constants except for the mobility. Once we obtain the mobility, we use Eq. 1 replacing $\Delta I_D$ and $\Delta n_s$ with the dark current, $I_D$, and the dark sheet charge density, $n_s$, respectively, to obtain $n_s$:

$$n_s = \frac{I_D L}{V_{DS} q \mu W} \qquad (4)$$

To excite electron-hole pairs in the AlGaN layer, one has to illuminate with photons of energy exceeding the AlGaN bandgap. If the AlGaN layer were thick, all photons would be absorbed in the AlGaN and none would reach the GaN. However in the typical GaN HEMT, the AlGaN is ~ 20 nm, which is over an order of magnitude smaller than the reciprocal of the absorption coefficient.[18] Hence, photons do reach and get absorbed in the GaN layer, as is commonly observed in optical responses of the structure.[19] To eliminate the contribution of the GaN layer to the photocurrent and to the photovoltage, we need to subtract the photo-responses of the GaN bandgap instead of subtracting the dark current and dark voltage. This means that we need to illuminate at two wavelengths, one – slightly below the AlGaN bandgap, and the other – above it. We then subtract the two photocurrent responses to obtain the net AlGaN photocurrent and subtract the two photovoltage responses to obtain the net AlGaN photovoltage.

To use the method, only two photon energies are actually required. However, for an appropriate selection of the energies, we scanned the photon energies over the range between 3.7 and 4.4 eV to obtain spectra of the channel photocurrent and the surface photovoltage. In both the photocurrent and photovoltage spectra, the photo-response of the AlGaN was observed to start about 0.2 eV below the bandgap due to electro-absorption (the Franz-Keldysh effect)[20] and the values at that photon energy ($E_g - 0.2eV$) were used as the reference "dark" values.

The $Al_{0.25}Ga_{0.75}N(20nm)$/Fe-doped GaN(2µm) structure was grown on SiC (CREE Inc.). Mesas were dry etched in chlorine-based plasma. After removal of the GaN cap layer, 100 nm of $Si_3N_4$ was deposited by plasma enhanced chemical vapor deposition (PECVD). Contact pads and 3 µm wide gate trenches were dry etched in $Si_3N_4$. All the metal contacts were deposited using e-beam thermal evaporation. The source and drain Ohmic contact layer sequence was Ti(30nm)\Al(70nm)\Ni(30nm)\Au(100nm) annealed at 900 °C for 1 min in nitrogen ambient. The gate contact was a Ni(30nm)/Au(100nm) Schottky barrier. Spectral illumination and electrical measurement were carried out inside a dark Faraday box at atmospheric room temperature conditions. For illumination, we used a 300 Watt Xe light source. The light was monochromitized using a Newport Corporation double MS257 monochromator followed by order-sorting long-pass filters. During spectral acquisition, a constant voltage of 0.1 V was applied between the source and drain contacts. Current measurements were carried out using two Keithley 2400 source-meters. Surface photovoltage was measured using a Kelvin probe (Besocke Delta Phi GmbH). Details of the method have been given elsewhere.[7] Photovoltage and photocurrent were obtained in the same setup under identical illumination conditions.

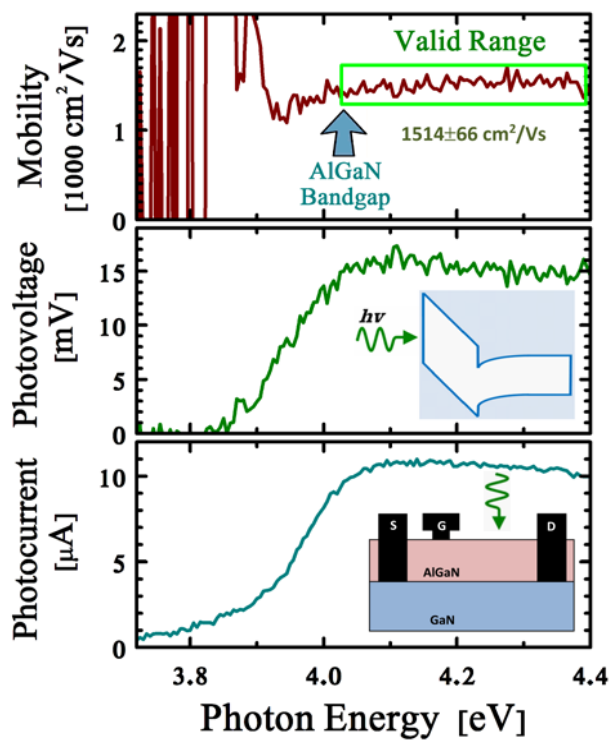

**Fig. 2** – Photocurrent (bottom panel), photovoltage (middle panel), and the resulting electron mobility spectrum (top panel) in AlGaN/GaN HEMT structure. Insets show the device structure (bottom) and the corresponding band diagram (middle panel) indicating the direction of the illumination.

Figure 2 shows the photocurrent (bottom panel), photovoltage (middle panel), and the resulting electron mobility spectrum (top panel) in AlGaN/GaN HEMT structure. It shows that the obtained mobility varies around an average value. These variations appear to be the result of the signal-to-noise ratio in the photovoltage spectrum. In the range preceding the AlGaN bandgap, the noise is amplified to extreme values due to a division by (nearly) zero. However, following the AlGaN bandgap the value becomes constant and is valid for mobility assessment. Averaging the values above the AlGaN bandgap and using the standard deviation as the measurement error, we obtained an electron mobility of ***1514±66 $cm^2/V·s$***. Once we have the mobility, the 2DEG sheet-charge density, $n_s$, may be readily obtained from the dark value of the current. We calculated a value of ***6.94±0.31·$10^{12}$ $cm^{-2}$*** for a dark current value of 1.68 mA. The two insets show the device structure and band diagram indicating the direction of illumination. For comparison, Hall effect measurements carried out on van der Pauw structures yielded ***1545±27 $cm^2/V·s$*** within the experimental error of the optical transconductance results.

In the case of the more common Hall effect method, Hall voltage is used to obtain the carrier concentration, and then the carrier concentration is used to obtain the mobility from a conductivity measurement. In the proposed method as well, we measure conductivity (channel current) but decompose it using surface photovoltage instead of the Hall voltage. The results demonstrate that the method affords a reliable alternative to Hall effect that allows to measure many transistors on a wafer to obtain statistics and wafer mapping.

## Acknowledgement

This work was funded by a research grant from the Israeli Ministry of Defense (MAFAT).